\documentclass[10pt]{article}
\usepackage{aaspp4}
\usepackage{lscape}     

\righthead{EXTRAPLANAR DUST IN EDGE-ON SPIRALS}
\lefthead{Howk \& Savage}

\newcommand{\HI}{H$\;${\small\rm I}\relax}
\newcommand{\HII}{H$\;${\small\rm II}\relax}
\newcommand{\Ha}{H$\alpha$\relax}
\newcommand{\Nh}{$N_H$\relax}

\newcommand{\z}{$z$}
\newcommand{\pI}{Paper~I}
\newcommand{\msun}{M$_\odot$}
\newcommand{\lsun}{$L_\odot$}
\newcommand{\etal}{{\em et al.}}
\newcommand{\column}{cm$^{-2}$}
\newcommand{\percc}{cm$^{-3}$}
\renewcommand{\v}{V}
\renewcommand{\b}{B}

\newcommand{\e}[1]{10^{#1}}

\newcommand{\av}{$a_V$}
\newcommand{\ab}{$a_B$}
\newcommand{\Av}{$A_V$}

\newcommand{\kms}{km~s$^{-1}$\relax}
\newcommand{\hst}{{\em HST}}

\begin{document}

%\submitted{To appear in {\em The Astronomical Journal}.}

\title{A Search for Extraplanar Dust in Nearby Edge-On
Spirals\altaffilmark{1}}

\altaffiltext{1}{Based on observations obtained at the WIYN
Observatory, a joint facility of the University of Wisconsin-Madison,
Indiana University, Yale University, and the National Optical
Astronomy Observatories.}

\author{J. Christopher Howk \& Blair D. Savage}
\affil{Department of Astronomy, University of Wisconsin-Madison, 
        Madison, Wi. 53706 \\ Electronic mail:
     howk@astro.wisc.edu, savage@astro.wisc.edu}

%\altaffiltext{2}{Current address: Department of Physics \& Astronomy,
%The Johns Hopkins University, 3400 N. Charles St., Baltimore, MD 21218}

\authoremail{howk@uwast.astro.wisc.edu}

\begin{center}
To appear in {\em The Astronomical Journal}, May 1999.
\end{center}

\begin{abstract}
        
We present high resolution ($0\farcs6$ to $\sim1\farcs0$) BV images of
12 edge-on spiral galaxies observed with the WIYN 3.5-m telescope.
These images were obtained to search for extraplanar ($|z| > 0.4$ kpc)
absorbing dust structures similar to those previously found in NGC~891
(Howk \& Savage 1997).  Many of these galaxies have been previously
searched for diffuse ionized gas at high \z.  Our imaged galaxies
include a sample of seven massive $L_*$-like spiral galaxies within
$D\la25$ Mpc that have inclinations $i\ga87^\circ$ from the plane of
the sky.  We find that five of these seven systems show extraplanar
dust, visible as highly-structured absorbing clouds against the
background stellar light of the galaxies.  These dust structures lie
at heights $|z| \ga0.4$ kpc, which should be above most of the thin
disk molecular material in these galaxies.  The more prominent
structures are estimated to have associated gas masses $\ga \e{5}$
M$_\odot$; the implied potential energies are $\ga \e{52}$ ergs.  {\em
All of the galaxies in our sample that show detectable \Ha\ emission
at large \z\ also show extraplanar dust structures.  None of those
galaxies for which extraplanar \Ha\ searches were negative show
evidence for extensive high-\z\ dust.  The existence of extraplanar
dust is a common property of massive spiral galaxies.}  We discuss
several mechanisms for shaping the observed dust features.  We
emphasize in this discussion the possibility that these dusty clouds
represent the dense phase of a multiphase medium at high-\z\ in spiral
galaxies.  In a few cases interactions with close galaxy companions
could be responsible for the high-\z\ dust, either through dynamical
stripping or triggered star formation.  We can rule out warps as the
source of the observed high-\z\ dust.  Flaring gas layers seem an
unlikely source of the observed material but cannot be ruled out at
this time, except for those features that clearly connect to energetic
processes in the disk.  The correlation between high-\z\ dust and
extraplanar \Ha\ emission may simply suggest that both trace the
high-\z\ interstellar medium in its various forms (or phases), the
existence of which may ultimately be driven by vigorous star formation
in the underlying disk.  The absorption produced by high-\z\ dust and
associated gas in spiral galaxies must be accounted for when studying
extraplanar emission from spiral galaxies over much of the
electromagnetic spectrum.
\end{abstract}
\section{INTRODUCTION}
\label{sec:intro}

Dust grains play an important role in the physics of the interstellar
medium (ISM) in the thin disks of galaxies, affecting the heating and
cooling of the gas, the phase structure of the medium, the
distribution of metals, and the transfer of radiation.  While the
processes that are thought to circulate matter between the thin disks
and halos of spiral galaxies will operate on both gas {\em and} dust,
the dust content of interstellar thick disks and halos is poorly
understood.  In our own Galaxy there is evidence from gas-phase
abundances that individual \HI\ clouds at relatively large distances
from the plane contain dust (Sembach \& Savage 1996, and references
therein).  However, these results give us little information on the
distribution of dust with distance \z\ from the Galactic midplane.

In an earlier paper we presented high-resolution optical images of the
nearby galaxy NGC~891 taken with the WIYN 3.5-m telescope (Howk \&
Savage 1997, hereafter \pI).  These images reveal extensive amounts of
highly-structured dust seen in absorption against the background
stellar light of the galaxy.  These dust structures, observable to
heights $0.4 \la z \la 2.0$ kpc from the midplane, are seen along the
entire length of the galaxy covered by the images presented in \pI.
We tentatively associated this dust absorption with the thick disk of
interstellar material observed in other wavebands.  An overly simple
analysis of the radiative transfer, coupled with gas-to-dust
relationships appropriate for the disk of the Milky Way (Bohlin,
Savage, \& Drake 1978), suggests these dust structures are relatively
opaque ($A_{V} \approx 0.8$ to 2.0) with correspondingly large gas
column densities ($N_{\rm H} \sim 10^{21}$ atoms \column) and masses
($M \sim 10^5$ to 10$^6$ \msun).  Estimates for the potential energies
of the observed dusty clouds relative to the midplane are in the range
$\Omega \sim \e{52} - \e{53}$ ergs, similar to the energy estimates
derived for Galactic supershells (Heiles 1979).  The implied total gas
mass associated with the ensemble of dusty high-\z\ clouds is roughly
similar to that estimated for the mass of the extraplanar diffuse
ionized gas (DIG) in NGC~891 ($\sim\e{8}$ M$_\odot$; Dettmar 1990).
While there are uncertainties in these order of magnitude estimates,
the images presented in \pI\ clearly show that a substantial amount of
dust is present in the thick disk of NGC~891.

How common is this phenomenon of extraplanar dust? Is it a peculiar
feature of one galaxy, or is there evidence in the local universe that
thick disk dust is a common feature of spiral galaxies?  In this paper
we try to partially answer these questions.  We have undertaken a
small imaging survey of edge-on galaxies within $D \la 25$ Mpc using
the WIYN 3.5-m telescope.  Our sample is limited to
northern-hemisphere galaxies with inclinations $i \ga 87^\circ$ from
the plane of the sky, though after obtaining our images, some of these
galaxies seemed less highly inclined.  Table \ref{table:galaxies}
gives a summary of the properties of our imaged galaxies.  The
properties listed in Table \ref{table:galaxies} are taken from
NED\footnote{The NASA/IPAC Extragalactic Database (NED) is operated by
the Jet Propulsion Laboratory, California Institute of Technology,
under contract with the National Aeronautics and Space
Administration.}, except for the distances which are from Tully
(1988)\footnote{The distance for NGC~4634 is not from Tully (1988).
We adopt a distance $d = 19$ Mpc from Teerikorpi \etal\ (1992)}.  For
each of these galaxies (with one exception) we have obtained 900s \b\
and \v\ images.  Not all of the imaged galaxies are appropriate for
the present work given the aforementioned inclination problems.  We
have exluded such galaxies from our ``final sample.''  We also limit
this final sample to only massive $L_*$ galaxies to avoid problems
with large changes in the gas-to-dust ratios.  We will show in this
paper that the presence of highly-structured dust at large heights ($z
\ga 400$ pc) from the midplane is a relatively common phenomenon and
seems to be correlated with the presence of extraplanar DIG.

In \S \ref{sec:observations} we give the details of our observations
and reductions; \S \ref{sec:unsharpmask} discusses our image display,
particularly our approach to producing unsharp masks to more clearly
reveal dust structures.  Our primary observational results are given
in \S \ref{sec:results}.  In \S \ref{subsec:individual} we briefly
discuss important properties of each of our target galaxies and give
reasons for exluding particular galaxies from our final sample.  A
discussion of the implications for the origins of high-\z\ dust and
for other observations of edge-on spirals is given in \S
\ref{sec:discussion}, and we summarize our main conclusions in \S
\ref{sec:summary}.

\section{OBSERVATIONS AND REDUCTIONS}
\label{sec:observations}

All of the observations presented here were obtained with the WIYN
3.5-m telescope at Kitt Peak National Observatory.  The WIYN imager in
use at the time of our observations is a thinned 2048$\times$2048 STIS
CCD with 21 $\mu$m pixels.  Placed at the f/6.5 Nasmyth focus of the
WIYN telescope each pixel corresponds to $0\farcs196$ on the sky.  The
imager has a $6\farcm7 \times 6\farcm7$ field of view.  The images
presented here were taken under non-photometric conditions, though the
seeing conditions were generally good.

A log of our observations for each galaxy is given in Table
\ref{table:log}.  This table shows for each image the filter used, the
date of observation, and the seeing-limited resolution for that
exposure, expressed as the FWHM (in arcseconds) of Gaussian fits to
the stellar images using the IRAF\footnote{IRAF is distributed by the
National Optical Astronomy Observatories, which is operated by the
Association for Research in Astronomy, Inc., under cooperative
agreement with the National Science Foundation.} routine {\tt IMEXAM}.
Also given is the linear resolution of our observations at the assumed
distance of each galaxy (see Table \ref{table:galaxies}).  With one
exception this survey is based upon 900s exposures of each galaxy in
the Harris \b\ and \v\ filters available at WIYN.  For the galaxy
NGC~4013 we include only a 600s V-band exposure.

The images have been bias-subtracted and flat field-corrected in the
usual manner within IRAF.  The flat fields were derived from
observations of the ``Great White Spot'' in the telescope dome.  We do
not have multiple exposures with which to correct for cosmic-ray
contamination, with the exception of the V-band data for NGC~4565, but
this does not seriously affect our results.

In principle subtle variations in the flat-field of the image could
mimic dust absorption in our target galaxies.  However, the vast
majority of the dust features we observe cannot be referred to as
``subtle.''  The imprint of small scale features in the CCD
flat-fields is relatively constant within a given night, allowing us
to search for features common to all galaxies that might be instrument
related.  Our experience suggests that the small-scale features
present in the flat-field are usually not significant compared with
the structures we are studying and are well removed in even single
exposures.  For our deeper imaging of a few galaxies (Howk \& Savage
1999a,b), we have taken exposures with the target galaxies shifted to
different parts of the imager to avoid complications associated with
flat-fielding errors.  Comparing the final images with the individual
frames, we are able to assess the probability of confusing flat-field
structures with dust structures within these galaxies; we have found
no real evidence for confusion of flat-field features with absorbing
dust structures.

\section{IMAGE DISPLAY AND UNSHARP MASK PROCEDURE}
\label{sec:unsharpmask}

Figures \ref{fig:n891}--\ref{fig:n5907} present the final results of
our imaging work.  For each galaxy we show the V-band image and an
unsharp masked version of the V-band data.\footnote{We choose to
display the V-band rather than B-band data because the higher signal
to noise ratio of our V-band images more than makes up for the greater
opacity expected from the dust features in the B-band.}  Since the
main results of our survey depend on our ability to effectively
display our images, it is worth discussing in detail our approach.  We
depend on unsharp masks of the V-band data to show the wealth of
structures that are present in many of these galaxies.  A good example
of the advantages to unsharp masks is Figure \ref{fig:n891}, which
shows our images of NGC~891.  The V-band image (bottom panel) is
displayed to show the distribution of light along the entire
length of the galaxy.  With this display the light from the bulge is
saturated and shows little in the way of distinguishable dust
features.  However, the unsharp mask of the V-band data (top panel)
allows us to show the absorbing structures along the entire length of
the galaxy in one display.  This display more accurately shows the
complexity of the absorbing structures that thread through the bulge
area.

The unsharp mask of an image was derived by dividing the original
image by a version smoothed with a Gaussian kernel.  The FWHM of the
Gaussian used to produce the smoothed image was 15 pixels ($3\farcs0$)
for galaxies closer than 10 Mpc and 7.5 pixels ($1\farcs5$) for
galaxies at greater distances (see Table \ref{table:galaxies}).  With
this approach we remove the large-scale gradients in the background
light of the galaxy, allowing us to display all parts of the
galaxy in a uniform manner.  The procedure tends to accentuate
structures on scales smaller than smoothing kernel ($\Delta \la
120-150$ pc for our galaxies).  This can be both an advantage and
disadvantage.  Most of the structures we find in these galaxies have
minor axis scales similar or smaller to this, though not all (see \S
\ref{subsec:features}).  In some sense this offsets the natural
tendency to have one's eye drawn towards very large structures.
Although we tend to identify structures only in computer displays of
the science images (i.e., not the unsharp masks), the reader's eye may
be drawn to the more small-scale structure in our unsharp masks.

Bright stars near a target galaxy can produce large artifacts in
unsharp masks of images, destroying information over a
disproportionately large area.  We have produced our unsharp masks by
dividing the original data by a smoothed image with such problem stars
replaced by a two dimensional fit to the surrounding background light
before the smoothing process.  In a few cases we have also removed
bright cosmic rays in the same manner.  The replaced area is usually
chosen to be a circular aperture of several times the FWHM of the
seeing disk.  For fainter examples, the aperture radius was typically
twice the size of the seeing disk, while the brighter stars required
apertures often in excess of five times the FWHM of the seeing disk.
We have also removed the effects of CCD blooming from a few bright
stars (e.g., Figures \ref{fig:n891} and \ref{fig:n4217} for NGC~891 and
NGC~4217, respectively).  This extra processing step to reduce the
effects of bright stars has been limited to stars lying on or near the
target galaxies.  In a few cases foreground stars lay in front of the
bright light of a galaxy bulge; we have avoided removing stars in
cases where the background to be fit comes from rapidly varying bulge
light.

By dividing the original image by the smoothed image with bright stars
and other features removed, we are able to keep bright stars from
influencing large regions of the target galaxy, while at the same time
showing where these stars lay with respect to the target.  The latter
point is important, because artifacts can be present in our unsharp
masked images very near the position of a bright star.  One can see
examples of the detrimental effects of even relatively faint sources
in Figure \ref{fig:n4217}.  In this image the stars well away from the
galaxy have not been removed in the masking process, and therefore the
unsharp mask process introduces the low-intensity halos surrounding
many of the stars (and background galaxies).  Even relatively faint
stars in our image of NGC~4217 exhibit extended halos in the unsharp
mask of our V-band image.  However one can see in this image that the
very bright star just north of the galaxy that was removed before the
smoothing process does not show a large-scale halo.  There are
artifacts present near the edges of this star, but it does not
overwhelm the galaxy as it does in unsharp masks produced without
removing it from the image.

In general one should be wary of subtle features near point sources or
even bright extended objects in our unsharp masks.  A lesson to this
effect is the image of NGC~4565 shown in Figure \ref{fig:n4565}.  This
galaxy shows a relatively prominent bulge (more so than most of the
galaxies in our sample).  The unsharp mask of the V-band image shows a
faint halo coincident with the relatively sharp edge of the bulge
light seen in the V-band image.  Faint, large-scale artifacts (white
in our unsharp mask) can be seen where the bulge of the galaxy is
brightest.  In particular, an artifact is present just to the south of
the dust lane where the bright center of the galaxy, whose emission
changes on small length scales, dominates the light.  Within the bulge
area itself there are point sources (presumably foreground stars) that
are hidden by the light of the bulge in our V-band data.  These have
not been masked out, and seem to exhibit slight artifacts.  NGC~4565
is a good example of artifact production in our unsharp mask
procedure, since the galaxy has very little high-\z\ dust.  We have
been careful to avoid producing artifacts.  This is part of the reason
we use different size smoothing boxes for galaxies depending on their
distances: we have found that this scale of smoothing produces the
most reliable results.

While there are possible problems associated with the interpretation
of the unsharp masks displayed here, one can see that this approach is
much more effective at showing the reader the true extent and
complexity of the dust features we see at high-\z\ in many of these
galaxies.

\section{RESULTS}
\label{sec:results}
	\subsection{Extraplanar Dust in the Survey} 

It is immediately clear from Figures \ref{fig:n891}--\ref{fig:n5907}
that many of the galaxies in our sample show evidence for dust above
the central dust lane.  Only dust that is highly structured or
significantly overdense relative to its surroundings is expected to be
obvious in our images.  Thus there could be a smoothly-distributed
component of dust at high-\z\ in all of these galaxies.  For our
purposes, we are interested in dust structures at heights $z \ga 400$
pc, i.e., well above the central absorbing dust lanes in these
galaxies.  For comparison the FWHM of the thin disk molecular gas is
estimated to be $\sim220$ pc in NGC~891 (Scoville \etal\ 1993), $\sim
150$ pc in NGC~4013 (Garc\'{\i}a-Burillo, Combes, \& Neri 1999, but
adopting our assumed distance of 17 Mpc), and $\sim100$ pc for the
Milky Way (e.g., Scoville \& Sanders 1987).  Thus structures with
heights $z \ga 400$ pc should lie above most of the material
associated with the thin interstellar disks of these galaxies.  In
\pI\ it was pointed out that the optical identification of dust
features such as those seen in our images is susceptible to a number
of selection effects.  We are much more sensitive to absorbing
structures that lie preferentially on the near side of a galaxy.  In
\pI\ we characterized dust structures at high-\z\ in NGC~891 by their
``apparent extinction,'' $a_\lambda$.  We defined the apparent
extinction in a waveband $\lambda$ as
\begin{equation}
 a_\lambda = -2.5 \log( S_{dc, \lambda} / S_{bg, \lambda} ),
\label{eqn:extinct}
\end{equation}
where $S_{dc, \lambda}$ is the surface brightness measured towards a
dust cloud, and $S_{bg, \lambda}$ is the surface brightness of the
local background.  The light measured towards a given dust feature
contains an extincted component of starlight originating behind the
feature along the line of sight as well as an unextincted component
emitted in front of the feature.  The apparent extinction in the
V-band, $a_V$, is a lower limit to the true extinction, $A_V$.  An
absorbing structure that has a very large $A_V$ with a large amount of
unextincted stellar light lying between it and the observer can have a
very small $a_V$ and show too little contrast to be detectable.  We
cannot identify high-\z\ dust in galaxies with very thin stellar light
distributions since we depend upon the background light of the stellar
distribution to illuminate the dust in which we are interested.

Table \ref{table:dust} summarizes our observational results, listing
our observational sample in order of decreasing star formation rate
(SFR) per unit area of the visible disk.  This quantity is traced by
the far-infrared (FIR) luminosity, $L_{FIR}$, divided by $D_{25}^2$,
the square of the major axis diameter defined by a surface brightness
contour of 25 mag. arcsec$^{-2}$ (e.g., Rand 1996).  The FIR
luminosities are derived from the {\em IRAS} 60 \micron\ and 100
\micron\ fluxes (Fullmer \& Lonsdale 1989).  This tracer of the SFR of
a galaxy is imperfect, however, given the contribution of the general
interstellar radiation field to the heating of dust grains (see Rand
1996 for a discussion of this point).  In particular the fraction of
the FIR luminosity that is a result of heating by the interstellar
radiation field may be a function of Hubble type (see Roberts \& Haynes
1994, and references therein).

Those galaxies that we have deemed appropriate to be included in our
``final sample,'' i.e., massive $L_*$-like galaxies that are
highly-enough inclined, are identified in Table \ref{table:dust}.  In
practice we do not use the observed visual luminosity of our target
galaxies to determine whether or not they should be included in this
final sample of luminous spirals given the possibly very important
effects of extinction in these edge-on systems.  Instead we choose to
use the observed \HI\ velocity profiles of our targets as an indicator
of their mass, and by extension their luminosity.  Table
\ref{table:galaxies} lists the velocity full-width at $20\%$ peak,
$W_{20}$, derived from the integrated \HI\ profile for each of our
target galaxies.  These data are taken from the {\em Third Reference
Catalogue of Bright Galaxies} (or RC3; de Vaucouleurs \etal\ 1991).
We exclude galaxies from our sample if their rotation velocities
(e.g., $W_{20}/2$) are less than 200 \kms\ at the $3\sigma$ level.
Where possible we have used the detailed rotation curves of galaxies
rather than the integrated \HI\ profiles of the galaxies from the RC3
in making this judgement.  We comment in \S \ref{subsec:individual} on
the reason(s) for excluding an imaged galaxy from this final sample,
giving references for the detailed rotation curves where appropriate.

Various well-studied edge-on galaxies have not made it into the
current survey for one reason or another.  Well-known galaxies have
often been excluded because they are too distant (e.g., NGC~5746,
NGC~5775, and UGC~10288), have inclinations $i<87^\circ$ (e.g.,
NGC~3079, NGC~3556, and NGC~5023), are not spiral galaxies (e.g.,
NGC~4762), or have rotation velocities suggesting they are not massive
$L_*$-like galaxies (e.g., NGC~3432, NGC~4244, IC~2233, and M~82).
Our selection by inclination is quite subjective in most cases.  For
galaxies that were close to the cut-off we have usually excluded them
from this sample on the basis of short WIYN images.

For each galaxy we make a judgement: is there evidence for significant
extraplanar dust or not?  Table \ref{table:dust} identifies those
galaxies in our survey that show extraplanar dust with a {\Large
$\bullet$} symbol, and those that show little or no evidence for
high-\z\ dust with a {\Large $\circ$} symbol.  A few cases were
ambiguous given their inclinations (e.g., NGC~4157 and NGC~4183).
Also given in Table \ref{table:dust} is an indicator of the presence
or absence of high-\z\ DIG, where a search has been made, and the
general morphology of the DIG where present.  References to the
appropriate DIG studies are given in the table.  An immediate
conclusion that can be drawn from Table \ref{table:dust} is that every
galaxy in our sample that was previously known to have high-\z\ DIG
shows evidence for extraplanar dust and none of those lacking
observable high-\z\ DIG shows any evidence for extraplanar dust.

A large fraction of the galaxies in our sample shows evidence for
high-\z\ dust.  Two cautionary comments are in order regarding our
sample.  The simpler of the two is that our sample size is quite
small.  We include seven galaxies in our final sample, five of which
show extraplanar dust.  The statistical uncertainties in a sample of
this size are expected to be quite large.  Assuming a binomial
probability distribution we find a formal result of $(71\pm17)\%$ of
spiral galaxies in the local universe have highly-structured
extraplanar dust.  However, our second caution is that our sample, in
all likelihood, is neither statistically complete nor unbiased.  We
have subjectively judged the inclination of the galaxies which have
poorly-known inclinations and chosen only those that visually appear
to have inclinations $i \ga 87^\circ$.  We cannot guarantee we have
included all appropriate galaxies, though most large galaxies in the
local universe (outside of the zone of avoidance) are probably known.
To our knowledge our final sample contains all of the {\em truly}
edge-on $L_*$-like northern spirals within 25 Mpc.  However, we are
likely biased to large, well-studied galaxies, particularly those that
have previously been searched for high-\z\ DIG.  Another possibly
important aspect of Table \ref{table:dust} is the non-uniformity of
the observations of high-\z\ DIG.  Even the individual galaxies within
the larger surveys of Rand (1996) and Pildis, Bregman, \& Schombert
(1994) have varying sensitivities to \Ha\ emission.

Though the searches for high-\z\ DIG are not statistically rigorous, a
picture has been developed based upon observations of more than 25
galaxies that couples the presence of DIG at high-\z\ with the SFR of
the underlying disk.  And though not many galaxies show bright,
widespread diffuse \Ha\ emission at high-\z, like NGC~891, many of the
galaxies studied (in the range of 60-70\%) show {\em some} form of
high-\z\ DIG (e.g., Rand 1998; Dettmar 1998). However, the
observations of this sample of galaxies are inhomogeneous.  If we
assume the presence of dust and ionized gas at high-\z\ in spiral
galaxies is tied to the SFR, perhaps a more meaningful number might be
the number of spiral galaxies with SFRs similar to those in our
sample.  

Galaxies with SFRs similar to NGC~4013 and NGC~891 are likely
relatively common.  For example, the Milky Way has a value of
$L_{FIR}/D^2 \sim 3\times\e{40}$ ergs s$^{-1}$ kpc$^{-2}$ (Cox \&
Mezger 1989; see Rand 1996), which is midway between the values of
this quantity for NGC~891 and NGC~4013.  The total $L_{FIR}$ of the
Galaxy is estimated to be $\sim 4\times \e{43}$ ergs s$^{-1}$, or
$\sim \e{10}$ \lsun\ (Cox \& Mezger 1989).  Roberts \& Haynes (1994)
have compiled a volume-limited sample of galaxies in the Local
Supercluster culled from the RC3 for comparing the properties of
galaxies along the Hubble sequence.  In their sample of Sab,Sb
galaxies they derive a median FIR luminosity of $L_{FIR} = [ 3.7 \,
(1.3, 7.9)] \times \e{9}$ \lsun, where the numbers in parentheses give
the the 25th and 75th percentile ranges.  This median is based on a
sample of 171 galaxies.  Their sample of 214 Sbc,Sc galaxies gives a
median $L_{FIR} = [4.8 \, (1.5, 11.5)] \times \e{9}$ \lsun.  The
median value of $L_{FIR}$ for both our full and final samples is
$L_{FIR} = 3.7\times \e{9}$ \lsun\ (using the upper limits as
detections).  These values are comparable to the FIR luminosity
distribution in the Roberts \& Haynes sample, suggesting that our
small sample of galaxies is not unusual compared with a larger sample
in the Local Supercluster, at least as judged by the median values of
$L_{FIR}$.  Further, the FIR luminosities of the individual galaxies
in our sample are not extraordinary given the broad distribution of
$L_{FIR}$ in the Roberts \& Haynes sample.  Though $L_{FIR}$ is an
imperfect measure, this suggests the SFRs in our target galaxies may
also typical of galaxies in the local universe.

The presence of dust at large distances from the midplane can no
longer be viewed as a peculiar property of a single galaxy (e.g.,
NGC~891).  Our observations reveal that the existence of
inhomogeneously distributed (clumped) high-\z\ dust is a common
property of $L_*$-like spiral galaxies.

	\subsection{Notes on Individual Galaxies}
	\label{subsec:individual}

We comment here briefly on the important properties of the individual
galaxies in our sample.  In particular we point out those galaxies
that are not included in our final sample and describe why they were
left out.  We also comment on galaxies known to be interacting or in
close groups, as well as those for which we have obtained more
extensive WIYN imaging that will be presented in future works.

{\em NGC 891 --} This galaxy has long been known to harbor extraplanar
dust (e.g., Sandage 1961; Keppel \etal\ 1991) and has been extensively
studied at many wavelengths (see \pI\ and references therein).  We
have obtained deeper \b\v I and \Ha\ images for this galaxy and will
present these results elsewhere (Howk \& Savage 1999a).  NGC~891 is
known to have an extensive distribution of DIG at high-\z\ (Dettmar
1990; Rand, Kulkarni, \& Hester 1990; Dettmar \& Schulz 1992; Rand
1996, 1997).  There is no evidence for a warp and the evidence for an
flare in the \HI\ layer has recently been called into question
(Sancisi \& Allen 1979; Swaters, Sancisi, \& van der Hulst 1997).
Alton \etal\ (1998) have recently presented submillimeter ($\lambda
850 \, \mu$m) images of NGC~891; they seem to detect emission from
dust up to $z \sim 2$ kpc from the midplane, confirming our results of
\pI.

{\em NGC 3628 --} This well-studied galaxy is part of the Leo Triplet,
which includes the galaxies NGC~3627 and NGC~3623.  It is experiencing
a close interaction with these nearby galaxies.  The western edge of
NGC~3628 shows a strong asymmetrical flare or warp seen in both \HI\
and optical images (e.g., Wilding, Alexander, \& Green 1993).  The
observed extraplanar dust in this system may be the result of the
strong interaction with its neighbors, either through gravitational
stripping, extreme flaring, or induced star formation.  However,
Wilding \etal\ comment that within the central 4\arcmin\ of the galaxy
the \HI\ disk shows very little high-\z\ emission and is close to
unresolved at their 15\arcsec\ resolution.  This galaxy houses a
circumnuclear starburst and has an extended hot gaseous halo (Dahlem
\etal\ 1996).

{\em NGC 4013 --} This galaxy shows extensive high-\z\ dust in our
WIYN images.  We have obtained deeper \b\v I and \Ha\ images of this
galaxy, the results of which will be presented elsewhere (Howk \&
Savage 1999b).  Rand (1996) has found faint diffuse \Ha\ emission
originating from high-\z\ DIG in this galaxy.  Garc\'{\i}a-Burillo
\etal\ (1999) have recently reported on high-resolution ($\sim3
\arcsec$) CO observations of this galaxy.  Several of the high-\z\
filaments in their channel maps correspond to dust structures seen in
our images.  While a detailed comparison of the CO emission and the
structures seen in our images is beyond the scope of our paper, we
note that a dust feature\footnote{This particular feature is
identified as ``Cloud 4'' in Howk \& Savage 1999c.} projected against
the bulge of the galaxy, stretching to $z \sim 750$ pc, is coincident
with an extraplanar CO feature in the maps of Garc\'{\i}a-Burillo
\etal\ (1999).  The velocity of this feature is $-104$ \kms\ with
respect to the systemic velocity of the galaxy.  This feature is
almost directly above the dynamical center of the galaxy; if it were
participating in the circular rotation of the galaxy (as in a flare),
the velocity relative to systemic should be nearly zero.  We believe
this argues against this particular feature being a result of a flared
gas layer at large distances from the center of the galaxy.  NGC~4013
is known to have a prodigious \HI\ warp perpendicular to the line of
sight (line of nodes parallel to the line of sight), and possibly also
a flared \HI\ layer (Bottema 1995, 1996).  The onset of the \HI\ warp
coincides with the end of the optical disk (Bottema 1995).  We do not
believe warps can explain the high-\z\ dust structures we observe in
our target galaxies; we will discuss this issue in \S
\ref{subsec:warps}.

{\em NGC 4157 --} The inclination of this galaxy makes the
identification of true high-\z\ dust ambiguous given the geometric
projection effects.  We do not include it in our final sample but 
present the images here for the reader's benefit.  

{\em NGC 4183 --} This galaxy may suffer from inclination effects as
for NGC~4157.  Also the rotational velocity of this galaxy suggests
that it may not be $L_*$-like in mass (see Rhee \& van Albada 1996).
We do not include this galaxy in our final sample.

{\em NGC 4217 --} This galaxy shows extensive high-\z\ dust along most
of its length, including one spectacular loop to the SW of the galaxy
center that can be traced to $z \sim 1.5$ kpc (NGC 4217:D $-016+015$
in Table \ref{table:features} and Figure \ref{fig:features2}).  We
have obtained deeper \b\v I and \Ha\ images of this galaxy, the
results of which will be presented elsewhere (Howk \& Savage 1999b).
Rand (1996) has identified two faint patches of \Ha\ emission
originating from high-\z\ DIG in this galaxy.  The estimated SFR of
this galaxy, derived from its $L_{\rm FIR}$, is quite low.  Thus it
does not fit the rough correlation observed between the presence of
high-\z\ DIG (as well as dust) and SFR (see Rand 1996 and Dettmar
1998).

{\em NGC 4302 --} Large scale loops and filaments of dust are seen at
high-\z\ along most of the length of this galaxy.  It forms a close
pair on the sky with NGC~4298.  NGC~4302 itself shows little evidence
for interaction, although NGC~4298 appears slightly distorted in the
direction of NGC~4302.  The DIG of this galaxy has been studied by
Rand (1996) who finds a relatively widespread, diffuse layer of \Ha\
emission detectable to $z\approx2$ kpc.  The RC3 gives $W_{20} = 377
\pm 8$ for this galaxy.  Thus while its measured rotation velocity is
less than our cut-off of $W_{20}/2 > 200$ \kms, it is consistent with
this value at the $3\sigma$ level.  Following Roberts \& Haynes (1994)
we derive a total mass for this galaxy of $\sim9.6\times \e{10}$
\msun, which is quite close to the other massive galaxies included in
our final sample, most of which weigh in with $\sim 1-3\times \e{11}$
\msun.  We therefore include NGC~4302 in our final sample.

{\em NGC 4517 --} This galaxy appears to be similar to NGC~4565 in its
inclination and high-\z\ dust content.  However, it has a less
prominent bulge, which makes the identification of high-\z\ dust more
difficult.  We find a few structures that may be dust at large \z, but
like NGC~4565 we find little evidence for widespread thick disk dust
like that seen in NGC~891.  Given the ambiguity caused by the inclination
of this galaxy, and the lack of light against which to view dust
absorption, we exclude this galaxy from our final sample.

{\em NGC 4565 --} This galaxy shows very little evidence for
widespread extraplanar dust like that seen in NGC~891 and several
other galaxies in our sample.  However, there are two rather prominent
extensions of dust seen to the NW of the bulge in Figure
\ref{fig:n4565}.  One of these features is identified in Table
\ref{table:features} and Figure \ref{fig:features3} (see \S
\ref{subsec:features}) as NGC 4565:D $-064-016$.  Given the lack of
similarity between this galaxy and the others showing high-\z\ dust,
we classify it as a galaxy without significant amounts of high-\z\
dust in Table \ref{table:dust}.  NGC~4565 shows evidence for both an
optical (Naeslund \& Joersaeter 1997) and \HI\ warp (Rupen 1991).
Though there is no evidence for \Ha\ or radio continuum emitting halos
(Rand, Kulkarni, \& Hester 1992; Sukumar \& Allen 1991; Dumke \etal\
1995), this galaxy does show high-\z\ X-ray emission (Vogler, Pietsch,
\& Kahabka 1996).  Given the absence of high-\z\ dust, this galaxy
provides useful information about artifacts introduced by the unsharp
masking process.

{\em NGC 4631 --} This galaxy is experiencing a close interaction with
the nearby galaxies NGC~4656 and NGC~4627.  The observed extraplanar
dust in this system may be caused by a mechanism different than that
responsible for high-\z\ dust in the more isolated galaxies of our
sample, such as tidal stripping or other dynamical effects.
Observations of \HI\ emission in this system (Weliachew, Sancisi, \&
Gu\'{e}lin 1978) show large amounts of gas between NGC~4631 and
NGC~4656, presumably stripped from one or both; these observations
also indicate that the \HI\ disk of NGC~4631 is warped or disturbed.
This galaxy is not included in our final sample because it rotates
more slowly than one expects for a massive $L_*$-like galaxy (see
Weliachew \etal\ 1978).  High-\z\ DIG in this galaxy has been imaged
by Rand, Kulkarni, \& Hester (1992) and studied spectroscopically by
Golla, Dettmar, \& Domg\"{o}rgen (1996).  NGC~4631 is also known to
have a very extended radio continuum halo with vertically oriented
magnetic fields (Dumke \etal\ 1995; Hummel, Beck, \& Dahlem 1991).

{\em NGC 4634 --} We find this galaxy has an extended thick disk or
halo of emitting material (presumably stars) that seems to flare with
increasing projected distance from the center.  Near the center of the
galaxy, this extended light distribution seems to have a projected
half-thickness of $\ga 2$ kpc, assuming a distance of 19 Mpc
(Teerikorpi \etal\ 1992).  Dettmar (1998) has presented an \Ha\ image
of this galaxy showing a bright DIG layer.  This galaxy is not
included in our final sample because of its low rotational velocity.

{\em NGC 5907 --} This well-studied Sc galaxy is characterized by a
relatively low SFR and has no detectable high-\z\ dust or ionized gas
(Rand 1996).  NGC~5907 shows a warp in the optical (Morrison \etal\
1994; S\'{a}nchez-Saavedra, Battaner, \& Florido 1990) and in \HI\
emission (Sancisi 1976; Shang \etal\ 1998).

	\subsection{Properties of Individual Dust Features}
	\label{subsec:features}

In Figures \ref{fig:features1} through \ref{fig:features3} we show
close-up views of some of the more prominent individual dust
structures in our target galaxies.  The properties of each feature
shown in Figures \ref{fig:features1}--\ref{fig:features3} are
summarized in Table \ref{table:features}.  This table lists the ID of
each feature as well as its position in J2000 equatorial coordinates.
These positions have been determined with respect to stars common to
our images and the Digitized Sky Survey and should be accurate to
$\sim1\arcsec$.  Throughout this section we will use a common naming
scheme to identify these individual features.  We designate a
structure using the form NGC GGGG:D $\pm$XXX$\pm$ZZZ.  Here the NGC
number of the individual galaxy is given as GGGG (clouds in UGC
galaxies should have a parallel naming scheme UGC GGGG:D
$\pm$XXX$\pm$ZZZ), and the ``D'' denotes a dust cloud.  The value XXX
is the projected distance in arcsec of the dust feature from the
optical center of the galaxy traced along the major axis.  Positive
values are east of the center.  The value ZZZ is the projected
distance in arcsec of the dust feature from the midplane of the galaxy
traced along the minor axis.  Positive values are north of the
midplane.  Applying this naming scheme to feature 2 from \pI\
[$(\alpha, \delta)_{\rm B1950} = (2^h\ 19^m \ 20.4^s, \ 42^\circ$
06\arcmin\ 44\arcsec)] we have NGC 0891:D $-044+032$.  The position of
the midplane has roughly been determined for this purpose using the
distribution of galaxy light as a function of height above the plane.
We have determined the point of symmetry in the absorption of the
central dust lane in cuts perpendicular to the galaxy plane in each of
our images, and we adopt this reference point in specifying ZZZ.  The
center of the galaxy in the radial direction is similarly determined
using cuts parallel to the major axis.  These designations are only
meant to be approximate to give the reader a sense of their radial and
vertical positions in the galaxy, and the equatorial coordinates
should be used for making detailed comparisons with other wavebands.
In the case of NGC~4302, whose position angle is almost 0$^\circ$, we
define positive XXX values as towards the north, and positive ZZZ
towards the {\em west} to maintain a consistent orientation of the
coordinate system in the sense that the positive $z$-axis is clockwise
from the $x$-axis in our figures.

In Table \ref{table:features} we give approximate $z$-heights as well
as rough dimensions of each structure.  We also give the apparent
extinctions \ab\ and \av\ in the B and V bandpasses, respectively, as
defined in Eqn. (\ref{eqn:extinct}).  We have measured these values as
described in \pI.  The errors in these measurements are approximately
$0.05 - 0.10$ mag.  As we pointed out earlier, the apparent extinction
measured towards an object is not equal to the true extinction because
of foreground stellar light contamination and the effects of
scattering of background stellar light into the structure.  The
apparent extinction does, however, represent a lower limit to the true
extinction in a given bandpass.  With information in three colors one
can attempt to roughly disentangle the amount of true extinction
(e.g., see \pI); however, applying this rough approach to only two
bandpasses seems dubious.  In Table \ref{table:features} we estimate a
lower limit to the hydrogen column density of associated gas for each
dust feature.  We assume Galactic gas to dust relationships and $R_V
\equiv A_V / E(B-V) = 3.1$, which is the average for diffuse clouds in
the Galactic disk (e.g., Cardelli, Clayton, \& Mathis 1989).  Using
the results of Bohlin \etal\ (1978) for the relationship of $N_{\rm
H}$ with $E(B-V)$ with our assumed $R_V$, we estimate the column
density as
\begin{equation}
N_{\rm H} >  1.9 \times 10^{21} \ a_V \ 
        {\rm [cm^{-2}]}.
\label{eqn:columndensity}
\end{equation}
We also then give a lower limit to the mass of each feature derived
from the product of this estimated column density and the projected
area of the feature.  The addition of light from stars residing in
front of the dust features is not the only effect that tends to lower
the apparent extinction from the true extinction.  Another possibly
significant effect is the unknown contribution of scattering off of
dust grains and into the line of sight.  This serves to lessen the
perceived extinction and make it more ``grey,'' i.e. less wavelength
dependent (see \pI).  Indeed, the effects of the ill-constrained
radiative transfer can be seen in Table \ref{table:features}.  If $R_V
\approx 3.1$ we expect $A_B = 1.34 A_V$.  If our dust feature lay
between the observer and all of the starlight of the galaxy, and
scattering were negligible, we would expect to also find $a_B = 1.34
a_V$.  The absorbing structure NGC 4013:D $+032-012$, for example,
shows $a_B \approx 1.1 a_V$.  This could be a result of the radiative
transfer or of an incorrect $R_V$.

As in \pI\ the structures we present here represent the more prominent
ones and likely arise preferentially on the near side of each galaxy.
The last column of Table \ref{table:features} gives a subjective
description of the morphology of each structure.  Note that the cloud
boundaries outlined in Figures \ref{fig:features1} and
\ref{fig:features2} are imprecise and very subjective.  The
identification of individual ``clouds'' in these galaxies is not only
challenging, but in many cases perhaps inappropriate.  The features we
show may be part of larger structures, or they may simply be the
result of overlapping structures along the line of sight.  However, it
was noted in \pI\ that structures identified in WIYN images of NGC~891
generally did not break up in an image taken with the {\em Hubble
Space Telescope} (\hst).  While the \hst\ image presented in \pI\
showed more substructure than was apparent in our WIYN images, it did
not offer immediate evidence for a separation of our identified dust
features into multiple structures.

Many of the structures have observed widths approximately twice our
seeing-limited resolution.  This suggests that the limited resolution
of our images may be affecting our measurements.  In these cases the
true width may be lower than that quoted in Table
\ref{table:features}.  In \pI\ we compared the widths derived from our
WIYN images and the archival \hst\ image.  The minor axis lengths
measured in both the WIYN and \hst\ images seemed to agree, however
the results are somewhat dependent upon the subjective definition of
the cloud boundaries.  The linear resolution of the images presented
in \pI\ (and the NGC~891 images presented here) are significantly better
than for most of our current dataset.  For the more distant galaxies
in our sample, the limited resolution of our images may be adversely
affecting our measurements.  This is true for our measurements of \ab\
and \av\ as well, in the sense that observed at higher resolution the
values of \ab\ and \av\ may be greater.

The values \Nh\ of gas associated with the identified dust features
shown in Table \ref{table:features}, which assume Galactic gas to dust
relationships (see \pI), are in the range $N_{\rm H} > 3 \times \e{20}
- \e{21}$ \column.  Of course our selection of features is biased
towards the more opaque structures (and hence higher column density),
but the true \Av\ of a given dust feature may be significantly greater
than the \av\ listed in Table \ref{table:features}.  The masses
derived for these structures are similar to those derived for dust
features in NGC~891 in \pI.  We include two dust absorbing structures
from NGC~891 in our current sample so that the reader may compare
these structures with those identified in our other galaxies.  We have
remeasured the properties of features 2 and 7 from \pI, or NGC 0891:D
$-044+032$ and NGC 0891:D $-012-030$, respectively, in the current
nomenclature, and present them with the other structures in Table
\ref{table:features} and Figure \ref{fig:features1}.  Our results
suggest that the dust features observed in the more extended dataset
presented here are similar to those discussed in \pI\ for NGC~891.
This statement is based more on impressions than a vigorous
comparison, which would require a less biased approach to identifying
such structures.

The potential energies of the identified structures are likely quite
large.  The structures identified in Table \ref{table:features} in
common with \pI, NGC 0891:D $-044+032$ and NGC 0891:D $-012-030$,
were estimated to have potential energies relative to the midplane
$\Omega \sim 3 \times \e{52}$ in that earlier work.  Though the true
potential energy depends sensitively on the variation of the
gravitational acceleration with $z$ for a given galaxy (and hence on
the mass scale height and midplane density), these features from
NGC~891 are not unusual compared with the other features identified in
Table \ref{table:features}.  This suggests that the potential energies
of these other structures may also be $\Omega \ga \e{52}$ ergs,
assuming the derived masses are appropriate.

One can see from Figures \ref{fig:features1}--\ref{fig:features3} that
there is a large morphological diversity within the population of dust
structures found at high-\z\ in a given galaxy.  Evidently the
high-\z\ ISM in these galaxies is non-uniform.  We are able to observe
these dusty clouds because they contain a greater density of dust (and
perhaps gas) than their surroundings.  There may very well be a more
smoothly distributed component of dust in these galaxies.  The
observed range of morphologies need not imply a range of mechanisms
for expelling material from the thin interstellar disks of these
galaxies.  The specific morphology of a given parcel of material will
depend sensitively on the details of its current and past
environments.

\section{DISCUSSION}
\label{sec:discussion}

We have shown through images obtained with the WIYN observatory that
dust can be found at large distances from the midplanes of many nearby
spiral galaxies.  In \pI\ we discuss several possible mechanisms for
shaping the dust seen far from the plane of NGC~891, all of which
required the ejection of dusty material from the thin disk of that
galaxy.  Promising mechanisms include hydrodynamic or
magnetohydrodynamic flows, such as fountain (Shapiro \& Field 1976;
Bregman 1980; Houck \& Bregman 1990; de Avillez, Berry, \& Kahn 1998)
and chimney (Norman \& Ikeuchi 1989; see also Heiles 1990) flows;
expulsion via radiation pressure on dust grains (e.g., Davies \etal\
1998; Ferrara 1993; Ferrara \etal\ 1991; Franco \etal\ 1991); and
flows driven by magnetic field instabilities (e.g, Parker 1966, 1992).
We refer the reader to \pI\ for a detailed discussion of these models
and how they might help to shape the observed high-\z\ dust
structures.

We concluded in \pI\ that star formation, and the resulting energy
input into the ISM through radiation, stellar winds, and supernova
explosions, must play a role in whatever process(es) one imagines is
responsible for the dusty thick disk of material in NGC~891.  That
conclusion is strengthened in the current work by the striking
correlation between the presence of high-\z\ dust and the
detectability of thick distributions of DIG in our sample of galaxies.
Extraplanar DIG in galaxies has been studied more thoroughly than the
dust structures we consider [cf., reviews by Dettmar (1998) and Rand
(1998), and references therein].  The main finding of these works is a
connection of the DIG emission with the SFR of the underlying disk.
With very few exceptions, DIG far from the plane is found only in
those disk galaxies showing large SFRs (traced again by the $L_{FIR}$
derived from {\em IRAS} observations).  The correlation of high-\z\
DIG and dust found in the current work suggests that both phenomena
must be ultimately tied to the same overall driving force, i.e., star
formation.  

\subsection{The Effects of Gaseous Warps and Flares}
\label{subsec:warps}

The observed dust structures cannot be due to gaseous warps in our
target galaxies.  We showed in \pI\ that the rough number of dust
features was similar for each side of NGC~891, which is not expected in
the case of a warp along the line of sight; a warp in the plane of the
sky is ruled out from \HI\ observations (e.g., Swaters \etal\ 1997).
Furthermore, the galaxies in our sample that do show prominent warps,
NGC~3628, NGC~4013, NGC~4565, and NGC~5907, include two galaxies that
show no evidence for substantial amounts of high-\z\ dust and two that
show extensive high-\z\ dust (see Table \ref{table:dust}).  The
influence of flared gas layers in our target galaxies is less easy to
dismiss, i.e., the observed high-\z\ dust could be due to flares.
However, we do not believe this is the case.  The morphologies of some
features are highly suggestive of hydrodynamical shaping by star
formation (see \pI), and the correlation presented here between
high-\z\ dust and ionized gas also suggests a connection with the SFR
of the disk.  Many of the galaxies containing DIG at high-\z\ show
either very patchy \Ha\ emission, presumably associated with locally
enhanced SFRs in the disk (e.g., above spiral arms), or thick disk DIG
layers that cut off sharply at a given projected distance from the
galaxy center (e.g., Rand 1996; Dettmar 1990; Rand \etal\ 1990).
Neither of these observational phenomena is explained if the DIG at
large projected distances from the midplane is due to a flared gas
layer.  We believe this argues against flares as an explanation for
the thick DIG layers of galaxies, and by extension against flares as
an explanation for high-\z\ dust.  The aforementioned coincidence of
dust features in NGC~4013 with extraplanar CO clouds
(Garc\'{\i}a-Burillo \etal\ 1999) at velocities incompatable with
galactic rotation further suggests that at least some of these dusty
clouds are not associated with a gaseous flare.

\subsection{A Multi-Phase ISM at High-\z}

Though the high-\z\ ionized gas and dust observed in spirals seem to
be highly correlated, we do not believe that they trace the same
material.  The clouds traced by dust, if they contain significant
amounts of gas, must be denser than that traced by the general DIG.
In \pI\ we suggest an approximate gas density regime for the
extraplanar dust features is $\sim1 - 10$ \percc, assuming Galactic
gas-to-dust relationships.  At $z \approx 1$ kpc in NGC~891 the expected
average density of electrons associated with the DIG is $\langle n_e
\rangle \sim 0.06$ \percc\ (Rand 1997), with the true local density
being perhaps $0.2-0.3$ depending on the filling factor.  Given our
lower limits to the implied column densities of the features listed in
Table \ref{table:features}, these structures would have to extend
$>500-1000$ pc along the line of sight to have densities as low as the
DIG.  In most cases this is a significantly larger extent than the
observed dimensions in the plane of the sky.  It seems unlikely that
the structures we observe extend that far along the line of sight.
Work to be presented in a future publication (Howk \& Savage 1999a)
suggests that not all of the absorbing dust features have associated
\Ha\ emission.  If these high-\z\ dust clouds have gas densities as
high as 1 \percc\ and are ionized, they would have very large \Ha\
emission measures compared with the surrouding DIG.  Thus the dust we
see in absorption may be a denser, neutral (and possibly molecular)
``phase'' of the ISM at high-\z\ than is traced through the \Ha\
emission.

The dust-bearing clouds seen at high-\z\ may represent a stable dense
phase within a multi-phase ISM at large distances from the plane
(e.g., Wolfire \etal\ 1995b; McKee \& Ostriker 1977).  Consideration
of the thermal equilibrium of a medium can under the right conditions
yield multiple stable equilibria.  Thus regions of vastly different
densities and temperatures can coexist in pressure equilibrium with
one another.  The range in parameter space (pressure and density) over
which a multiphase medium is stable depends on the detailed heating
and cooling processes of the medium (cf., Wolfire \etal\ 1995a).  In
particular the existence of a stable multiphase medium requires the
pressure of the medium be in an appropriate range.  At higher
pressures only the dense phase is the stable, while at low pressures
the low-density phase dominates.

For spiral galaxies having dusty thick disk gas with a requisite
pressure, the formation of highly-structured clouds, such as those
observed in our WIYN images, may be caused by a relaxation to the
stable thermal equilibrium state of the medium.  The presence of dust
grains is extremely important for determining the conditions of the
medium given the heating provided by photoelectron emission from dust
and the cooling through thermal FIR radiation.  Dust incorporates
elements such as C, O, Si and Fe that might otherwise provide
significant gas-phase coolants through IR fine structure emission
lines.

Wolfire \etal\ (1995a) discuss the minimum required pressures to
provide a thermally stable dense phase of the ISM in various
environments.  For a medium with standard Milky Way disk abundances,
radiation fields (far-UV and X-ray), and dust to gas ratios, Wolfire
\etal\  find a minimum pressure $P^{min}/k \approx 1000$ K \percc\ is
required for the existence of a dense phase, assuming a cloud with
$N_{\rm H} \approx \e{19}$ \column.  The column density of the cloud
is important because the heating and ionization rates due to soft
X-rays and far-UV radiation are sensitive to the absorbing column
density.  For clouds of order $N_{\rm H} \approx
\e{20}$ \column, closer to those observed in this work, they find
$P^{min}/k \approx 600$ K \percc.  Other factors are important for
such considerations, including the strength of the interstellar
radiation field and the gas to dust ratio; in the case of the high-\z\
material discussed here these factors all tend to decrease the minumum
pressure required for a stable dense gas phase, with the exception of
possible variations in the C and O abundances (see Wolfire \etal\
1995a).

Are the pressures at high-\z\ in our target galaxies sufficiently high
to support a stable dense phase of the ISM?  In the case of NGC~891,
the best studied of these galaxies, the answer is yes.  Observations
of X-ray and \Ha\ emission from NGC~891 have provided some information
on the electron density distribution at high-\z\ (Bregman \& Houck
1997 and Rand 1997, respectively).  Bregman \& Houck (1997)
characterize the X-ray emitting gas with a temperature $T\sim3.5\times
\e{6}$ K, central density $n_o \sim 0.026$ \percc, and a Gaussian
scale height perpendicular to the plane of $h_z \sim 3.5$ kpc.  The
radial scale-length of this gas is thought to be $h_R \sim 4.1$ kpc.
Thus the midplane pressure should be of order $P/k \sim \e{5}$ and
half that value at $R\approx 2$ kpc, $z\approx 1.0$ kpc.  Even at
$R\sim8$ kpc, comparable to the sun's distance from the Galactic
center, the expected pressure of the hot ISM according to the Bregman
\& Houck fits should be $P/k \sim \e{4}$ K \percc.  Thus if the
observed dusty clouds are immersed in the hot gaseous halo of this
galaxy at $z\sim1$ kpc, the pressures are much greater than those
required by the Wolfire \etal\ models.  We find similar results if the
clouds are embedded in the high-\z\ DIG.  At $z\sim1$ kpc the average
density of electrons in the thick ionized gas layer of NGC~891 should
be of order $\langle n_e \rangle \sim 0.06$ \percc, according to the
distribution derived by Rand (1997).  If we assume nebular
temperatures for this gas, in this case $T_e \sim 8000$ K, and an
ionization fraction $x_e \equiv n_e/n_{\rm H} \sim 0.8$ (Rand 1998),
then we find $P/k \sim 1100$ K \percc\ at $z\sim1$ kpc.  Again this is
sufficient to make the presence of dense gas in pressure equilibrium
at high-\z\ reasonable, though not as comfortably as for the hot gas.
If we assume a volume filling factor of $f\sim0.2$, making the local
electron density of the gas $n_e \sim 0.3$ \percc, the appropriate
pressure is more like $P/k \sim 5500$ K \percc.  Though we have not
discussed the effects of differing abundances, gas to dust ratios, and
radiation fields (which are discussed more fully for high-\z\ gas in
Wolfire \etal\ 1995b), the expected pressures at heights $z \sim 1$
kpc in NGC~891 seem sufficiently high for the presence of a stable
dense medium.  Note that non-thermal pressures may also be important.

Assuming the observed dust structures represent a dense phase of the
ISM at high-\z, one can estimate the thermal crossing times for these
structures.  This allows us to roughly estimate the expected lifetime
of a cloud assuming no sources of confinement.  Rough estimates of the
densities for our cloud structures, using our column density estimates
and minor axis size scales, suggest densities of order $n_{\rm H} \sim
1 - 10$ \percc.  At pressures characteristic of the DIG at high-\z,
e.g., $P/k \sim 1100 - 5500$ K \percc, this corresponds to
temperatures of order $T\sim 100 - 5500$ K (assuming pressure
equilibrium), with lower temperatures being more likely if these
structures truly represent a dense phase of the ISM.  If we assume
minor axis sizes of order 100 pc, the thermal crossing times for this
range of temperatures is of order $\e{7} - \e{8}$ years.  This is
similar to the expected circulation timescales for a parcel of gas
travelling from the disk to halo and back (e.g., Houck \& Bregman
1990; Norman \& Ikeuchi 1989).  The timescales may be longer than this
if magnetic fields are present.  This suggests that these structures
are not simply transient, but could indeed be long lived.

If the cool dust clouds are formed from thermal instabilities in the
hot or warm ionized medium at high-\z\, it implies that dust must
survive in the DIG and/or the hot gas.  This is not an unreasonable
scenario.  Howk \& Savage (1999d) have recently shown that dust must
exist in the warm ionized medium of the Milky Way based upon the
gas-phase abundances of Al and S.  And while there is little
observational information on the dust content of the hot ISM,
theoretical calculations suggest that grains are able to survive the
thermal sputtering in such a harsh environment for as much as $\e{9}$
years (Ferrara \etal\ 1991).

A scenario in which the observed high-\z\ dusty clouds represent the
dense phase of a multi-phase ISM at high-\z, like the models discussed
in \pI, requires a large star formation rate in the underlying disk to
provide both the material at high-\z\ and the requisite pressure.
Thus fountain- or chimney-like outflows may cause a region of
relatively high pressure just above the thin disk that can enable the
formation of dense cloud condensations in the range of heights where
we observe the absorbing structures considered in this work.  If
thermal equilibrium considerations drive the {\em in situ} formation
of the observed dusty clouds at high-\z, a natural consequence of the
large required pressure may be a large emissivity in the \Ha\ line
from the DIG.  The intensity of \Ha\ emission from a parcel of gas is
proportional to the product $n_p n_e$, or proportional to $n_e^2$ in a
pure hydrogen gas, where $n_p$ and $n_e$ are the proton and electron
densities, respectively.  If we assume the DIG exists within a range
of temperatures $T\sim 6000 - 10000$ K, the relatively large pressure
required to confine the dense clouds is manifested in a large density,
and hence fuels a large \Ha\ emissivity.  Thus galaxies that have
high-\z\ material with thick disk pressures large enough to support a
dense phase of the ISM may naturally be expected to show detectable
\Ha\ emission at high-\z.  

\subsubsection{High-\z\ Star Formation}

It is interesting that many of the dust features we see in our target
galaxies are of similar sizes and estimated masses to giant molecular
clouds in the disk of the Milky Way.  As we suggested earlier, the
observed dust features may indeed contain molecular material,
depending on the true gas to dust ratio and interstellar radiation
field at high-\z.  Given this similarity it is possible stars may be
presently forming from the high-\z\ material observed in many of our
galaxies.  Indeed, we have found evidence for high-\z\ ($0.6 \la z \la
1.0$ kpc) \HII\ regions in NGC~891 and NGC~4013 (Howk \& Savage
1999c).  These regions of ionized gas are presumably powered by
early-type stars, though spectroscopy is needed to confirm the
classification of these objects as \HII\ regions.  Such high-\z\ \HII\
regions have been noted by other authors for the galaxies NGC~4244
(Walterbos 1991) and NGC~55 (Ferguson, Wyse, \& Gallagher 1996).  It
would be worthwhile to carefully search for molecular emission (e.g.,
CO) at high-\z\ in the galaxies that do show these dust structures.

\subsection{Milky Way Analogs}

Are there structures in the Milky Way that are comparable to those
observed here?  This question is complicated by the difficulties of
identifying structures given our position in the disk of the Milky
Way.  Koo, Heiles, \& Reach (1992) have identified Galactic structures
they term ``worms'' seen in both \HI\ and {\em IRAS} maps of the
galaxy.  These dusty filaments may be similar to the dusty clouds we
see, though it is unclear if their vertical extents are as great as
the structures seen in our sample of galaxies. The Draco cloud, an
intermediate-velocity \HI\ and molecular cloud of great interest since
it absorbs background X-rays (Burrows \& Mendenhall 1991), has
recently been shown to lie at a height $300 \la z \la 400$ pc by
Gladders \etal\ (1998).  This cloud may be similar in column density
to the structures we discuss, but its mass and height above the plane
are significantly less than the structures we identify.  Callaway
\etal\ (1999) report on \HI\ observations of a Galactic supershell in
the inner galaxy (the ``Scutum Supershell'').  This structure
stretches to heights $z \la 600$ pc from the midplane and contains
some $\sim\e{5}$ \msun, with an estimated potential energy of $\Omega
\ga \e{52}$ ergs.  Furthermore, a high-\z\ cloud ($z\sim600$ pc)
containing $\sim4\times\e{4}$ \msun\ of material seems to be
associated with this structure.  The \HI\ column density of this
high-\z\ cloud is of order $3\times \e{20}$ \column, and its diameter
is $\sim170$ pc.  A structure such as the Scutum Supershell, or
perhaps the high-\z\ cloud associated with it, would be visible in our
galaxies, provided there was little stellar light along the line of
sight arising in front it.  Another possible Galactic analog may be
the Low-Latitude Intermediate Velocity Arch (LLIV Arch; Kuntz \& Danly
1996).  Wakker \etal\ (1996) have shown this structure to lie at a
height $1.1 \la z \la 3.0$ kpc from the plane of our Galaxy.  With
peak column densities of a few times $\e{20}$ \column, and a total
mass $\ga \e{5} - \e{6}$ \msun\ (Kuntz \& Danly 1996; Wakker \etal\
1996), this feature might be detectable if placed in our target
galaxies.  Given its relatively low column density (and its unknown
dust content) it might only be detectable if it were in front of most
of the starlight.

\subsection{High-\z\ Opacity Distribution}

One implication of the observed dust structures in many of our target
galaxies is that dust opacity cannot be neglected at heights $z \la
1.5$ kpc from the midplane.  In \pI\ we estimated values of $0.8 \la
A_{\rm V} \la 1.8$ mag. for a sample of absorbing structures at
heights $0.4 \la z \la 1.5$ kpc in NGC~891.  We have found similar values
for absorbing structures in NGC~4013 (e.g., Howk \& Savage 1999c).
Intercepting only one of these features along the line of sight
provides an optical depth $\tau_{\rm V} \sim 1$ (to light arising
behind it) in the visual wavebands.  The complicated and widespread
distribution of absorbing structures in many of these galaxies
suggests that an observer may expect to intercept more than one of
these features along sightlines having heights up to $z \sim 1$ kpc,
and perhaps further.  The true form of the \z -distribution of dust
opacity in edge-on galaxies may be poorly constrained by many of the
current fits to the vertical light distribution.  The presence of a
dusty thick disk of material with relatively large scale height (e.g.,
$\sim 1$ kpc) can create problems for fits to the one- or
two-dimensional light profiles of spiral galaxies.  The opacity
distribution caused by thin+thick disk distributions of dust can be
relatively well-fit by a single component dust model (K. Wood,
private communication), but the resulting scale height estimate of the
opacity is larger (smaller) than that of the true thin (thick) disk
opacity.  Thus previous estimates for the dust scale heights in
galaxies that show evidence for significant high-z opacity may need to
be questioned. 

If the gas-to-dust ratio in these dust structures is similar to that
found in the Galactic disk, the opacity to X-ray emission could also
be quite large.  For example the optical depths at 0.25, 0.5 and 1.0
keV associated with neutral disk material having $\tau_V = 1.0$ in our
Galaxy are $\tau (0.25, \ 0.5, \ 1.0 \ {\rm keV}) \approx (8.0, \ 1.2,
\ {\rm and} \ 0.5)$, respectively (B. Benjamin, private
communication).  However, the distribution of opacity along a given
sightline through a dusty thick disk of material is likely quite
patchy, making the transfer of radiation more difficult to account for
when attempting a correction for absorption.  Studies of the
distribution of material traced by X-ray, ultraviolet, or optical
light should take care to examine how the presence of dust having
optical depth unity at visual wavelengths to heights $z\sim1$ kpc
affect their results.

\section{SUMMARY}
\label{sec:summary}

We present the results of a high resolution ($0\farcs6$ to
$\sim1\farcs0$) imaging survey, using the WIYN 3.5-m telescope, of
nearby edge-on galaxies, the purpose of which is to search for
high-\z\ dust in absorption against the background stellar light of
our target galaxies.  This work follows from the identification of
hundreds of high-\z\ absorbing structures seen in the galaxy NGC~891 by
Howk \& Savage (1997).  A summary of our major results is as follows.

\begin{enumerate}

\item Inhomogeneously distributed (clumped) high-\z\ dust is a common
property of $L_*$ spiral galaxies; five of seven such galaxies in our
final sample exhibit high-\z\ ($z \ga 400$ pc) dust structures.

\item There is a close correlation between the presence of high-\z\
DIG and dust in our sample.  Of those galaxies that have been searched
for high-\z\ \Ha\ emission (nine total, seven from our final sample),
all that show such emission also show observable dust structures at
high-\z\ (seven of nine, or five of seven drawn from our final
sample), none of those galaxies lacking \Ha\ emission at high-\z\ show
evidence for dusty material far above the disk (two of nine, or two of
seven from our final sample).  This correlation may be due to a common
driver of high-\z\ material and is probably associated with vigorous
star formation in the underlying disk.

\item We derive the properties of several individual dust structures,
which show sizes of order $50-100\times100-400$ pc.  In general we
find apparent extinctions $a_V \sim 0.2 - 0.9$, though the true V-band
extinction may be significantly greater than these values.  Assuming
Galactic gas to dust ratios are appropriate, the masses of these
structures exceed $\sim\e{5}$ \msun.  The potential energies of these
structures are likely quite large, of order $\Omega \ga \e{52}$ ergs
relative to $z = 0$.  The dust features exhibit a wide range of
morphologies.  One of the more impressive examples is seen above the
bulge of NGC~4217 (NGC 4217:D $-016+015$); this structure appears as
a loop of diameter $\sim0.8$ kpc, maximum height $z \sim 1.3$ kpc, and
total mass $>5 \times \e{5}$ \msun.

\item We comment on the possible mechanisms responsible for creating
and shaping the observed dust structures.  Possible mechanisms include
hydrodynamical flows (e.g., galactic fountains or bores and chimney
models), flows assisted or driven by magnetic fields or radiation
pressure, and {\em in situ} formation of high-\z\ clouds from cooling
high-\z\ gas.  Three of our galaxies have close companions; in these
cases the high-\z\ gas and dust may be caused by galaxy interactions,
either through triggered star formation or tidal stripping.

\item We discuss in detail the possibility that the dusty clouds we
observe represent a dense phase of the high-\z\ ISM in pressure
equilibrium with the extraplanar warm ionized or hot medium.  Rough
calculations for the best-studied galaxy NGC~891 suggest that the
high-\z\ ISM in this galaxy satisfies the requisite conditions for
support of a dense phase (e.g., Wolfire \etal\ 1995a,b).  Our rough
estimates for the density of the observed structures are compatable
with the Wolfire \etal\ results for high column density clouds.  We
discuss the possibility that star formation may be occuring in these
dense high-\z\ clouds.

\end{enumerate}
\acknowledgements

It is a pleasure to thank the operators at WIYN who, through their
expertise, have made this survey possible: C. Corson, W. Hughes,
C. Mackey, and G. Rosenstein.  We also thank T. Tripp for sharing some
of his WIYN observing time.  JCH thanks M. de Avillez and R. Dettmar
for enlightening discussions that have helped this work.  This
research has made use of the NASA/IPAC Extragalactic Database (NED)
which is operated by the Jet Propulsion Laboratory, California
Institute of Technology, under contract with the National Aeronautics
and Space Administration.  Our work has also made use of the SIMBAD
database, operated at CDS, Strasbourg, France.  JCH recognizes support
from a NASA Graduate Student Researcher Fellowship under grant number
NGT-5-50121.

\clearpage
\pagebreak

\begin{planotable}{lcccccccc}
\tablenum{1} 
\tablewidth{0pc}
\tablecolumns{10}
\tablecaption{Edge-On Galaxy Targets\tablenotemark{a}
	\label{table:galaxies}}
\tablehead{\colhead{Name} & \colhead{RA} & \colhead{Dec.} &
\colhead{$D_{25}$} & \colhead{V$_{rad}$} & \colhead{$W_{20}$} &
\colhead{Dist.} & \colhead{$L_{FIR}$} & \colhead{Type} \\
	\colhead{} & \colhead{[J2000]} & \colhead{[J2000]} & 
	\colhead{[arcmin]} & \colhead{[km s$^{-1}$]} & 
	\colhead{[km s$^{-1}$]} & \colhead{[Mpc]} & 
	\colhead{[$10^9$ L$_\odot$]} & 	\colhead{} 	
}
\startdata
NGC 891   & 02 22 33 & +42 20 48 & 12.2 & 
	528  & 471 & 9.6 & 9.9    &  Sb  \nl
NGC 3628  & 11 20 16 & +13 35 22 & 14.8 & 
	843  & 476 & 7.7 & 5.2    &  Sb  \nl
NGC 4013  & 11 58 32 & +43 56 48 & 4.7  & 
	835  & 407 &  17  & 3.7    &  Sb  \nl 
NGC 4157  & 12 11 05 & +50 29 10 & 7.0  & 
	774  & 413 & 17  & 8.5    &  Sbc \nl
NGC 4183  & 12 13 18 & +43 41 55 & 5.1  & 
	930  & 249 & 17  & $<0.2$ &  Scd \nl
NGC 4217  & 12 15 51 & +47 05 32 & 5.1  & 
	1026 & 431 & 17  & $<0.2$ &  Sb  \nl 
NGC 4302  & 12 21 42 & +14 36 05 & 4.7  & 
	1149 & 377 & 17  & $\la3.3$ &  Sc  \nl
NGC 4517  & 12 32 28 & +00 23 23 & 9.5  & 
	1131 & 164 & 9.8 & 1.0    &  Sc  \nl 
NGC 4565  & 12 36 21 & +25 59 05 & 16.2 & 
	1220 & 525 & 9.7 & 1.6    &  Sb  \nl
NGC 4631  & 12 42 08 & +32 32 28 & 14.7 & 
	632  & 315 & 6.9 & 4.1    &  Sc  \nl
NGC 4634  & 12 42 40 & +14 17 47 & 2.6  & 
	297  & 280 & 19\tablenotemark{b} & 3.2   &  Sc  \nl 
NGC 5907  & 15 15 54 & +56 19 46 & 11.2 & 
	667  & 485 & 15  & 5.0    &  Sc  \nl
\enddata
\tablenotetext{a}{The properties presented here were taken from the
NED database with the exception of the distances, which were taken
from Tully (1988) and the far-infrared luminosities, $L_{FIR}$, which
were derived using the adopted distances and the far-infrared fluxes
of Fullmer \& Lonsdale (1989).  The width of the \ion{H}{1} profile at
20\% of the peak flux, $W_{20}$, is taken from the RC3 (de Vaucouleurs
{\em et al.} 1991).}
\tablenotetext{b}{The distance to NGC~4634 is taken from Teerikorpi
{\em et al.} (1992) and is based on a Tully-Fisher magnitude-line width
relationship. }
\end{planotable}

\pagebreak

\begin{planotable}{lcccc}
\tablenum{2} 
\tablewidth{0pc}
\tablecolumns{5}
\tablecaption{Log of Observations\tablenotemark{a}
\label{table:log}}
\tablehead{
\colhead{Galaxy} & \colhead{Filter} &
\colhead{Date}   & \colhead{Seeing} &
\colhead{Resolution\tablenotemark{b}} \nl
\colhead{} &\colhead{} &\colhead{} &
\colhead{[arcsec]} & 
\colhead{[pc]} }
\startdata
NGC 891  & B & 1997 Aug. 29 & 0.6 & 27 \nl %2099
         & V & 1997 Aug. 30 & 0.6 & 27 \nl %3073
NGC 3628 & B & 1998 Mar. 02 & 0.9 & 34 \nl %5063
         & V & 1998 Mar. 02 & 0.9 & 34 \nl
NGC 4013 & B & 1997 Apr. 08 & 0.6 & 49 \nl
         & V\tablenotemark{c}
             & 1997 Apr. 08 & 0.5 & 41 \nl
NGC 4157 & B & 1998 Apr. 17 & 0.9 & 74 \nl %1067
         & V & 1998 Apr. 16 & 1.0 & 82 \nl %0046
NGC 4183 & B & 1998 Apr. 17 & 1.0 & 82 \nl %1070
         & V & 1998 Apr. 16 & 1.0 & 82 \nl %0046?
NGC 4217 & B & 1998 Apr. 17 & 0.8 & 66 \nl %6050
         & V & 1998 Apr. 17 & 0.8 & 66 \nl %6050?
NGC 4302 & B & 1997 May. 12 & 0.9 & 74 \nl %0071
         & V & 1998 Apr. 17 & 0.8 & 66 \nl %1052
NGC 4517 & B & 1998 Apr. 17 & 0.9 & 41 \nl %1060/61
         & V & 1998 Apr. 17 & 1.1 & 50 \nl %1049/50
NGC 4565 & B & 1997 Apr. 07 & 0.8 & 38 \nl %0061
         & V\tablenotemark{d} 
             & 1996 Dec. 03 & 0.8 & 38 \nl %0062
         & V\tablenotemark{d} 
             & 1997 Apr. 07 & 1.0 & 53 \nl %0062
NGC 4631 & B & 1998 Apr. 17 & 0.8 & 27 \nl %1066
         & V & 1998 Apr. 16 & 1.0 & 33 \nl %0047
NGC 4634 & B & 1998 Apr. 17 & 1.0 & 19: \nl %1064
         & V & 1998 Apr. 17 & 0.8 & 16: \nl %1063
NGC 5907 & B & 1997 Apr. 06 & 1.3 & 95 \nl %0031
         & V & 1998 Apr. 17 & 1.5 & 110 \nl %1074?
\enddata
\tablenotetext{a}{All observations have 900s exposure times except
         where noted.}
\tablenotetext{b}{Linear resolution of the observations assuming the
 	distance given in Table 1.}
\tablenotetext{c}{The exposure time for the NGC 4013 V-band data
	presented here is 600s.}
\tablenotetext{d}{The NGC 4565 V-band data were split between two
	nights.  The 1996 Dec. data have a 300s exposure time; the
	1997 Apr. data have 600s exposure time.  We have coadded the
	data after smoothing the Dec. data to $1\farcs 0$ resolution.}
\end{planotable}

\pagebreak 

\newcommand{\wyes}{\Large $\bullet$ \normalsize}
\newcommand{\wno}{\Large $\circ$ \normalsize}
\newcommand{\noedge}{$\times$}
\newcommand{\yes}{$+$}
\newcommand{\no}{$-$}
\newcommand{\final}{$\surd$}

\begin{planotable}{lcccccl}
\tablenum{3} 
\tablewidth{0pc}
\tablecolumns{7}
\tablecaption{Dust and DIG Properties for Edge-On Galaxy Sample
\label{table:dust}}
\tablehead{\colhead{Name} & \colhead{Final} &
\colhead{L$_{FIR}$/D$^2_{25}$} & 
\colhead{High-z} & \colhead{High-z} & \colhead{DIG} &
\colhead{DIG}\\
  \colhead{} & \colhead{Sample\tablenotemark{a}} &
  \colhead{(10$^{40}$ erg/s kpc$^2$)} & 
  \colhead{Dust\tablenotemark{b}} & 
  \colhead{DIG\tablenotemark{c}} & 
  \colhead{Ref.} & 
  \colhead{Morphology\tablenotemark{d}}}
\startdata
NGC 4634 &        & 5.9  & \wyes & \wyes  & 1  & Very bright; extended \nl
NGC 891  & \final & 3.3  & \wyes & \wyes  & 2,3& Bright; diffuse+filaments \nl
NGC 4157 &        & 2.7  & \no   &        &    &                \nl
NGC 4013 & \final & 2.6  & \wyes & \wyes  & 4  & Faint; diffuse  \nl
NGC 4302 & \final & $\lesssim$2.3 
                         & \wyes & \wyes  & 4 & Faint; diffuse \nl
NGC 3628 & \final & 1.8  & \wyes & \wyes  & 5 & \nl
NGC 4631 &        & 1.8  & \wyes & \wyes  & 6 & Bright; diffuse \nl
NGC 5907 & \final & 0.8  & \wno  & \wno   & 4 & \nl
NGC 4517 &        & 0.5  & \wno  &        &   & \nl
NGC 4565 & \final & 0.3  & \wno  & \wno   & 6 & \nl
NGC 4217 & \final &  $<$0.12 
                         & \wyes & \wyes  & 4 & 2 faint patches \nl
NGC 4183 &     &  $<$0.12 
                         & \no   &    &   & \nl
\enddata
\tablenotetext{a}{Some of the galaxies we have imaged were excluded
 from our final sample due to low inclinations or low rotational
 velocities (see text).  We have marked galaxies included in our final
 sample with a \final\ mark.}
\tablenotetext{b}{\wyes\ denotes galaxies that exhibit high-$z$ dust
 in WIYN images; \wno\ denotes galaxies that do not exhibit high-$z$ dust
 in WIYN images.  A few cases were ambiguous due to inclination
 effects; these are marked with \no\ symbol.}
\tablenotetext{c}{\wyes denotes galaxies with observable high-$z$ DIG;
 \wno\ denotes galaxies for which H$\alpha$ searches have not shown
 detectable high-$z$ DIG.}
\tablenotetext{d}{After Dettmar (1998) and Rand (1996).}
\tablerefs{ (1) Dettmar (1998); (2) Dettmar (1990); (3) Rand,
Kulkarni, \& Hester (1990); (4) Rand (1996); (6) Fabbiano, Heckman, \&
Keel (1990); (5) Rand, Kulkarni, \& Hester (1992) }
\end{planotable}

\pagebreak

%\ptlandscape
\begin{landscape}
\begin{planotable}{lccccccccl}
\tablenum{4} 
\tablewidth{0pc}
\tablecolumns{10}
\tablecaption{Properties of Individual High-$z$ Dust Features
\label{table:features}}
\tablehead{
\colhead{ID\tablenotemark{a}} & 
\colhead{R.A.} & \colhead{Dec.} & 
\colhead{$z$\tablenotemark{b}} & 
\colhead{Dimensions} &
\colhead{$a_{B}$\tablenotemark{c}}  & 
\colhead{$a_{V}$\tablenotemark{c}}  & 
\colhead{$N_{\rm H}$\tablenotemark{d}}  & 
\colhead{Mass\tablenotemark{e}} & 
\colhead{Morphology} \nl
\colhead{}  &  \colhead{[J2000]} & \colhead{[J2000]} &
\colhead{[pc]} & \colhead{[pc$\times$pc]} &
\colhead{[mag.]}       &  \colhead{[mag.]}  &
\colhead{[cm$^{-2}$]}    &  \colhead{[M$_\odot$]} & 
\colhead{} 
}
\startdata
NGC 0891:D $-044+032$\tablenotemark{f}
	 & 02 22 29.1 & $+42$ 20 23
                 & 1450 & $50\times100$ & 0.85 & 0.60 &
        $>1\times\e{21}$ & $>1\times\e{5}$ &  Elongated Cloud \nl
NGC 0891:D $-012-030$\tablenotemark{f}
	& 02 22 35.2 &  $+42$ 20 29 
                 & 1350 & $60\times140$ & 0.44 & 0.32 &
        $>6\times\e{20}$ & $>1\times\e{5}$ &  Cometary Cloud \nl
NGC 3628:D $-030-032$ 
	& 11\ 20\ 15.0 & $+13$\ 34\ 42 
		& 1120 & $70\times160$ & 0.68 & 0.50 &
        $>1\times\e{21}$ & $>8\times\e{5}$ & Thin Sheet \nl
NGC 3628:D $-053+039$ 
	& 11\ 20\ 14.4 & $+13$\ 35\ 58 
		& 1450 & $75\times210$ & 0.28 & 0.25 &
        $>5\times\e{20}$ & $>1\times\e{5}$ & Irr. Cloud \nl
NGC 4013:D $+032-012$ 
	& 11\ 58\ 34.3 & $+43$\ 56\ 52
		& 1000 & $250\times300$ & 0.57 & 0.52 & 
	$>1\times\e{21}$ & $>9\times\e{5}$ & Irr. Cloud \nl
NGC 4013:D $-001-032$ 
	& 11\ 58\ 31.4 & $+43$\ 56\ 44
		& 640  & $100\times230$ & 0.46 & 0.35 &
	$>6\times\e{20}$ & $>2\times\e{5}$ & Irr. Cloud \nl
NGC 4217:D $-016+015$ 
	& 12\ 15\ 48.6 & $+47$\ 05\  32
		& 1260 & $130\times800$\tablenotemark{g}
				 & 0.21 & 0.18 &   
	$>3\times\e{20}$ & $>5\times\e{5}$  & Large Loop \nl
NGC 4302:D $-007+006$ 
	& 12\  21\  41.7 & $+14$\  35\  44 
		& 500  & $150\times350$ & 0.36 & 0.31 &
        $>6\times\e{20}$ & $>4\times\e{5}$  & Irr. Cloud \nl
NGC 4302:D $+029-018$ 
	& 12\  21\  43.5 & $+14$\  36\  18
		& 1450 & $130\times260$ & 0.60 & 0.45 & 
        $>8\times\e{20}$ & $>3\times\e{5}$ & Irr. Cloud \nl
NGC 4302:D $+038+009$ 
	& 12\ 21\  41.6 & $+14$\  36\  27
		& 710 & $130\times200$ & 0.68 & 0.54 & 
	$>9\times\e{20}$ & $>5\times\e{5}$ & Irr. Cloud \nl
NGC 4565:D $-064-016$ 
	& 12\ 36\ 17.0 & $+26$\ 00\ 00 
		& 700 & $210\times400$ & 1.20 & 0.90 &
        $>2\times\e{21}$ & $>1\times\e{6}$ & Vert. Column\nl
NGC 4634:D $-016-010$ 
	& 12\ 42\ 40.8 & $+14$\ 17\  33 
		& 1000 & $90\times180$ & 0.58 & 0.33 &
        $>6\times\e{20}$ & $>1\times\e{5}$ & Irr. Cloud \nl
\enddata

\tablenotetext{a}{Identification in the form NGC GGGG:D
$\pm$XXX$\pm$ZZZ, where GGGG is the four number galaxy identification,
XXX is the distance in arcsec from the optical center of the galaxy
traced along the major axis, and ZZZ is the distance in arcsec from
the optical center of the galaxy traced along the minor axis (see
text).  }
 
\tablenotetext{b}{Projected height above the midplane, or the limit to
which very extended features can be traced.}

\tablenotetext{c}{Apparent extinction for the BV wavebands in
magnitudes, as defined in the text.}  

\tablenotetext{d}{Approximate lower limit to the column density of
material assuming Galactic extinction and gas-to-dust conversions and
$R_{\rm V} = 3.1$.}

\tablenotetext{e}{Approximate mass based upon the estimated column
density and projected area.  Includes a factor of 1.37 correction for
He.}

\tablenotetext{f}{These two features in NGC~891 were discussed in
Paper~I as features 2 and 7 (NGC 0891:D $-044+032$ and NGC 0891:D
$-012-030$, respectively).  The potential energies of these structures
relative to the midplane of NGC~891 were estimated to be $\Omega \sim
3 \times 10^{52}$ ergs in Paper~I.}

\tablenotetext{g}{The dimensions given here represent the diameter of
the loop at maximum extent, and the thickness of the walls of the
loop.}

\end{planotable}
\end{landscape}

\pagebreak
\clearpage

\figcaption{WIYN V-band images of NGC 891.  The bottom panel shows
the V-band image of this galaxy, while the top panel shows an unsharp
masked version of the V-band data.  The box at the lower left corner
is 10\arcsec$\times$10\arcsec. At the distance of NGC~891, 10\arcsec\
corresponds to $\sim450$ pc.  North is to the top of the page and east
to the left.  \label{fig:n891}}

\figcaption{As Figure \ref{fig:n891}, but for the galaxy
NGC~3628.  At the distance of NGC~3628, 10\arcsec\
corresponds to $\sim375$ pc. \label{fig:3628} }

\figcaption{As Figure \ref{fig:n891}, but for the galaxy
NGC~4013.  At the distance of NGC~4013, 10\arcsec\
corresponds to $\sim825$ pc. \label{fig:n4013} }

\figcaption{As Figure \ref{fig:n891}, but for the galaxy
NGC~4157.  At the distance of NGC~4157, 10\arcsec\
corresponds to $\sim825$ pc.  This galaxy is not included in our final
sample.  \label{fig:n4157} }

\figcaption{As Figure \ref{fig:n891}, but for the galaxy NGC~4183.  In
this case the unsharp masked image is shown on the left and the V-band
on the right.  At the distance of NGC~4183, 10\arcsec\ corresponds to
$\sim825$ pc. This galaxy is not included in our final
sample.  \label{fig:n4183} }

\figcaption{As Figure \ref{fig:n891}, but for the galaxy
NGC~4217.  At the distance of NGC~4217, 10\arcsec\
corresponds to $\sim825$ pc. \label{fig:n4217} }

\figcaption{As Figure \ref{fig:n891}, but for the galaxy NGC~4302.  In
this case the unsharp masked image is shown on the left and the V-band
on the right.  At the distance of NGC~4302, 10\arcsec\ corresponds to
$\sim825$ pc. \label{fig:n4302} }

\figcaption{As Figure \ref{fig:n891}, but for the galaxy NGC~4517.  At
the distance of NGC~4517, 10\arcsec\ corresponds to $\sim450$ pc. This
galaxy is not included in our final sample. \label{fig:n4517} }

\figcaption{As Figure
\ref{fig:n891}, but for the galaxy NGC~4565. At the distance of
NGC~4565, 10\arcsec\ corresponds to $\sim450$ pc.  The unsharp mask
image of this galaxy shows a slight artifact surrounding the location
of the bright bulge of this galaxy.
\label{fig:n4565} }

\figcaption{As Figure \ref{fig:n891}, but for the galaxy NGC~4631.  At
the distance of NGC~4631, 10\arcsec\ corresponds to $\sim335$ pc. This
galaxy is not included in our final sample.  Many of the dust
structures in this galaxy seem to be larger than the adopted FWHM of
the smoothing Gaussian.  Therefore some of the features visible in the
V-band image (bottom panel) become less distinct in the unsharp mask
(top panel).
\label{fig:n4631} }

\figcaption{As Figure \ref{fig:n891}, but for the galaxy NGC~4634.  At
the distance of NGC~4634, 10\arcsec\ corresponds to $\sim920$ pc. This
galaxy is not included in our final sample.  The unsharp mask of this
galaxy shows faint artifacts running mostly parallel to the disk that
are caused by the extended thick stellar disk discussed in \S
\ref{subsec:individual}.
\label{fig:n4634} }
\clearpage

\figcaption{As Figure \ref{fig:n891}, but for the galaxy NGC~5907.  At
the distance of NGC~5907, 10\arcsec\ corresponds to $\sim730$
pc. \label{fig:n5907} }

\figcaption{Close-up views of individual dust features.  These
close-ups are taken from the unsharp-masked V-band images.  The ID for
each feature is given in a slightly compressed notation to save space,
and the properties of each feature are summarized in Table
\ref{table:features}.  We have lightly outlined each cloud.  In a few
cases we outline possible extensions of the cloud that were not used
for calculating the values of Table \ref{table:features} with dashed
lines.  \label{fig:features1}}

\figcaption{As Figure \ref{fig:features1} but showing different
structures.  \label{fig:features2}}

\figcaption{As Figure \ref{fig:features1} but showing different
structures.  \label{fig:features3}}

\end{document}